\documentclass[reprint,aps,prl,onecolumn,superscriptaddress,floatfix,longbibliography]{revtex4-1}

\usepackage[dvips]{graphicx}
\usepackage{latexsym}
\usepackage{amsmath}
\usepackage{amsthm}
\usepackage{amsfonts}
\usepackage{amssymb}
\usepackage{mathptmx}
\usepackage{subfigure}
\usepackage{mathrsfs}

\usepackage{color}
\definecolor{darkblue}{rgb}{0,0.02,0.45}
\usepackage[colorlinks=true,citecolor=darkblue]{hyperref}

\usepackage[all]{hypcap}
\usepackage{gensymb}
\definecolor{cream}{RGB}{222,217,201}

\usepackage[paper=letterpaper,twoside,top=2.0cm,bottom=2cm,left=2cm,
            right=2cm,bindingoffset=0.0cm,voffset=0.0cm]{geometry}
            
\usepackage{lipsum}
\makeatletter
\renewcommand\frontmatter@abstractwidth{\dimexpr\textwidth-0in\relax}
\makeatother

\begin{document}

\clearpage


\title{Reentrance of spin-driven ferroelectricity through rotational tunneling of ammonium}

\author{Yan Wu}
\thanks{These authors contributed equally: Yan Wu, Lei Ding and Na Su} 
\author{Lei Ding}
\thanks{These authors contributed equally: Yan Wu, Lei Ding and Na Su}
\affiliation{Neutron Science Division, Oak Ridge National Laboratory,
Oak Ridge, TN 37831, USA}
\author{Na Su}
\thanks{These authors contributed equally: Yan Wu, Lei Ding and Na Su} 
\affiliation{Beijing National Laboratory for Condensed Matter Physics and Institute of Physics, Chinese Academy of Sciences, Beijing 100190, China}
\affiliation{School of Physical Sciences, University of Chinese Academy of Sciences, Beijing 100190, China}

\author{Yinina Ma}
\affiliation{Beijing National Laboratory for Condensed Matter Physics and Institute of Physics, Chinese Academy of Sciences, Beijing 100190, China}
\affiliation{School of Physical Sciences, University of Chinese Academy of Sciences, Beijing 100190, China}

\author{Kun Zhai}
\affiliation{Beijing National Laboratory for Condensed Matter Physics and Institute of Physics, Chinese Academy of Sciences, Beijing 100190, China}
\affiliation{School of Physical Sciences, University of Chinese Academy of Sciences, Beijing 100190, China}

\author{Xiaojian Bai}
\author{Bryan C. Chakoumakos}
\affiliation{Neutron Science Division, Oak Ridge National Laboratory,
Oak Ridge, TN 37831, USA}

\author{Young Sun}
\affiliation{Beijing National Laboratory for Condensed Matter Physics and Institute of Physics, Chinese Academy of Sciences, Beijing 100190, China}
\affiliation{School of Physical Sciences, University of Chinese Academy of Sciences, Beijing 100190, China}
\affiliation{Songshan Lake Materials Laboratory, Dongguan, Guangdong 523808, China}

\author{Yongqiang Cheng}
\affiliation{Neutron Science Division, Oak Ridge National Laboratory,
Oak Ridge, TN 37831, USA}

\author{Jinguang Cheng}
\affiliation{Beijing National Laboratory for Condensed Matter Physics and Institute of Physics, Chinese Academy of Sciences, Beijing 100190, China}
\affiliation{School of Physical Sciences, University of Chinese Academy of Sciences, Beijing 100190, China}
\affiliation{Songshan Lake Materials Laboratory, Dongguan, Guangdong 523808, China}

\author{Wei Tian}
\affiliation{Neutron Science Division, Oak Ridge National Laboratory,
Oak Ridge, TN 37831, USA}

\author{Huibo Cao}
\email[]{caoh@ornl.gov}
\affiliation{Neutron Science Division, Oak Ridge National Laboratory,
Oak Ridge, TN 37831, USA}

\date{\today}

\baselineskip20pt
\begin{abstract}
\baselineskip20pt
\noindent\textbf{Quantum effects 
fundamentally engender exotic physical phenomena in macroscopic systems, which advance next-generation technological applications. Rotational tunneling that represents the quantum phenomenon of the librational motion of molecules is ubiquitous in hydrogen-contained materials.
However, its direct manifestation in realizing macroscopic physical properties is elusive. Here we report an observation of reentrant ferroelectricity under low pressure that is mediated by the rotational tunneling of ammonium ions in molecule-based (NH$_4$)$_2$FeCl$_5 \cdot$H$_2$O. Applying a small pressure leads to a transition from spin-driven ferroelectricity to paraelectricity coinciding with the stabilization of a collinear magnetic phase. Such a transition is attributed to the hydrogen bond fluctuations via the rotational tunneling of ammonium groups as supported by theoretical calculations. Higher pressure lifts the quantum fluctuations and leads to a reentrant ferroelectric phase concomitant with another incommensurate magnetic phase. These results demonstrate that the rotational tunneling emerges as a new route to control magnetic-related properties in soft magnets, opening avenues for designing multi-functional materials and realizing potential quantum control.}
\lipsum[0]
\end{abstract}

\maketitle

\lipsum[0]
Historically, magnetoelectric phenomenon which describes the cross-coupling between electricity and magnetism in solid was first conceived by Pierre Curie in a sense of symmetry breaking that basically creates the physical phenomena in physics \citep{Curie1894}.
Of particular interest are magnetoelectric multiferroics where the simultaneous presence of long-range magnetic and electric dipole orderings in the same phase enables the magnetic field control of electric polarization and vice versa \citep{Kimura2003,Lottermoser2004,Tokura2014,Spaldin2005,Khomskii2009, Kenzelmann2005,Hur2004,Bibes2008,Kun2017,Mandal2015}.
~A paradigmatic example is TbMnO$_3$ which was found to show gigantic spontaneous electric polarization due to the incommensurate spin ordering as a result of the innate spin frustration \citep{Kimura2003}. Spin frustration, as a consequence of exchange interaction competitions, is generally sensitive to even small geometrical modifications of crystal structure. Such modifications can be readily realized by hydrostatic pressurization, thus, providing a promising means to explore novel multiferroics and understand the underlying mechanisms \citep{Aoyama2014,Rocquefelte2013,Terada2014}. The application of a high pressure of 4.5 GPa on TbMnO$_3$ eliminates the spin frustration and drives the system from a cycloidal spin configuration into a collinear ferroelectric $E$-type magnetic structure which is persistent to the highest measured pressure of 8.7 GPa \cite{Aoyama2014,Terada2016}. However,in considering the requirement of high pressure, stabilizing a ferroelectric phase under relatively small pressures would be highly desired. 

\begin{figure*}
\centering
\includegraphics[width=1\textwidth]{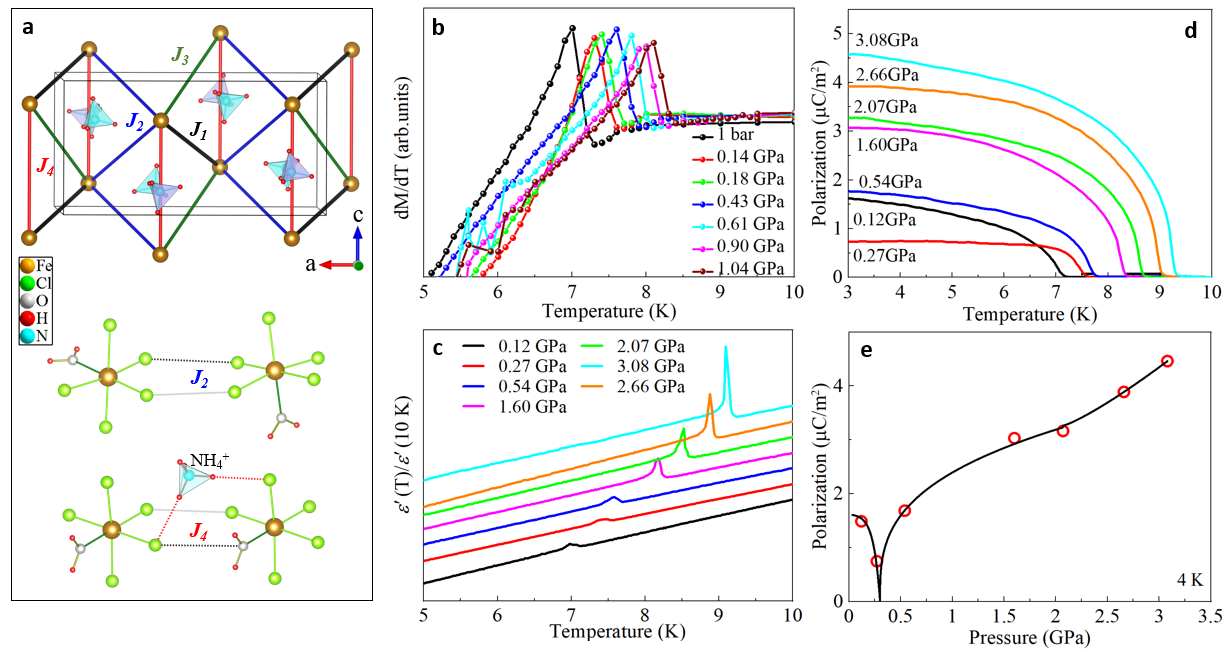}
\caption{\textbf{The crystal structure, magnetic, dielectric and ferroelectric properties of (NH$_4$)$_2$FeCl$_5 \cdot$H$_2$O}. \textbf{a}, Schematic drawing of the crystal structure and four possible exchange interactions between Fe cations in the $ac$ plane(space group $P2_1/a$). The ordered NH$_4^+$ groups are presented by the tetrahedra at two nitrogen sites: N1 (cyan tetrahedra) and N2 (purple tetrahedra). The dotted red line denote the strongest hydrogen bonds (Cl···H-N1) that mediate the exchange interaction $J_4$ between Fe ions along the $c$ axis. \textbf{b}, The derivative of the temperature-dependent magnetization under pressure with the magnetic field applied along the $c$ axis in the space group $Pnma$. \textbf{c}, The temperature dependence of the relative dielectric constant along the $a$ axis in the orthorhombic lattice under pressure. \textbf{d}, The temperature-dependent electric polarization along the $a$ axis in the orthorhombic lattice under pressure. \textbf{e}, The pressure-dependent electric polarization at 4 K. The solid curve is a guide to the eye to show the almost null polarization under 0.3 GPa.}\label{fig:1}
\end{figure*}

Molecule-based materials are tantalizing candidate materials. Owing to the presence of molecular groups and super-super-exchange interactions via directional hydrogen bonds, a delicate change in the lattice by a relatively small or moderate pressure may profoundly alter the magnetic and related properties \citep{Manson2009,Neal2016,Halder2011}. Moreover, molecular materials that contain hydrogen bonds entail inherent quantum fluctuations as extra freedom from hydrogen permutations apart from the spin fluctuations, which may facilitate the emergence of novel spin and electric states \citep{Prager1997,Geirhos2020,Horiuchi2008, Saparov2016,Horiuchi2003}. Here we present the pressure-induced ferroelectric reentrance in the molecular material (NH$_4$)$_2$FeCl$_5 \cdot$H$_2$O whose crystal structure and electric properties at ambient pressure are analogous to
the prototypical TbMnO$_3$ (Fig. \ref{fig:1}a) \citep{Ackermann2013}. By mapping out its magnetic and electric phase diagram under pressure, we present that the orientations of ammonium groups govern the magnetic phase transitions and results in the occurrence of ferroelectric reentrance. We argue that the consecutive transitions are likely steered by the quantum tunneling occurring between ammonium ions under pressure as supported by our theoretical calculations.

The temperature derivative of the magnetization curves of (NH$_4$)$_2$FeCl$_5 \cdot$H$_2$O with the applied magnetic field of 0.1 T along the $c$ axis of the orthorhombic lattice under pressure are shown in Fig. \ref{fig:1}b. The curve at 1 bar reveals a peak at 6.9 K, signaling the long range spin order, consistent with the critical temperature T$_{FE}$=6.9 K for the ferroelectric phase at ambient pressure \citep{Ackermann2013, Tian2016, Rodriguez2018}. The transition temperature monotonically increases with pressure likely due to the strengthened exchange couplings as a result of the compressed volume. No other anomaly at low temperatures was observed under pressure. Note the other magnetic transition around 7.3 K at ambient pressure is not evident in the magnetization curves, which is in agreement with prior work \citep{Ackermann2013}.
Figure \ref{fig:1}c shows the normalized dielectric constant $\epsilon'_a$(T)/$\epsilon'_a$(10 K) under pressures up to 3.08 GPa. A cusp was observed in the $\epsilon'_a$(T) at T$_{FE}$ = 6.9 K at ambient pressure, signaling the onset of the ferroelectric transition (FE I) \citep{Ackermann2013}. It becomes broader and weaker with the increase of pressure and practically vanishes around 0.3 GPa. With further increasing pressure, the sharp peak was unprecedentedly recovered and enhances considerably, suggesting a reentrant ferroelectric phase that correlates with the long range magnetic ordering as shown in the following. 
The electric polarization measurements show a finite spontaneous polarization along the $a$ axis that appears just around T$_{FE}$ under small pressures (Fig. \ref{fig:1}d). Figure \ref{fig:1}e shows the evolution of electric polarization as a function of pressure at 4 K. A sudden drop in the spontaneous polarization can be clearly discerned from 0 to 0.27 GPa, in good consistence with the dielectric constant results. With further increasing pressure, the electric polarization increases proportionally and reaches a value of 4.6 $\mu$C/m$^2$ at 4 K and 3.08 GPa, indicating that the system enters another pressure-induced ferroelectric phase (FE II in Fig. \ref{fig:3}c). These results show that the spin-driven ferroelectric phases, FE I and FE II, emerge in the low and high pressure regions, respectively. The electric polarization at 0.3 GPa looks minimal. Based on these results and the neutron diffraction data thereafter, we evidence that there appears a critical pressure around P$_C$ = 0.3 GPa where the spontaneous polarization vanishes and the system passes into a paraelectric (PE II) phase.   

\begin{figure}
\centering
\includegraphics[width=0.8\linewidth]{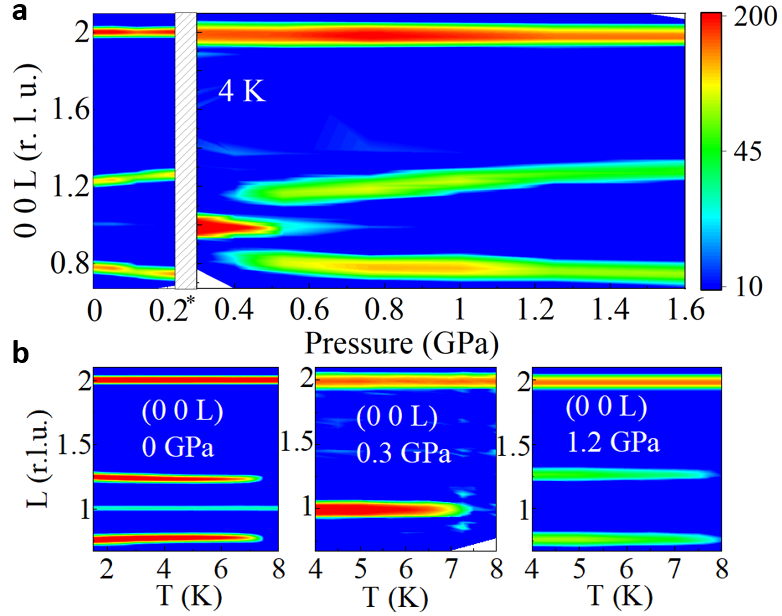}
\caption{\textbf{Contour plot of the neutron diffraction along (0 0 L) direction at different pressures and temperatures for (ND$_4$)$_2$FeCl$_5 \cdot$D$_2$O}. \textbf{a}, The evolution of the neutron diffraction data along the (0 0 L) direction as a function of pressure at 4 K. The asterisk denotes the pressure values applied by the Harwood gas pressure cell, see more details for pressure calibrations in the Extended Date file. The weak trace of (0 0 1) peak below 0.2$^*$ GPa is likely due to multiple scattering. \textbf{b}, The temperature-dependent neutron diffraction data at various pressures.}\label{fig:2}
\end{figure}

To unveil the underlying mechanisms of the ferroelectric phase transitions, we have performed neutron diffraction measurements on deuterated single crystals under pressure. Figure \ref{fig:2}a shows the contour map of neutron diffraction along the (0 0 L) direction collected under various pressures at 4 K. In coincidence with the dielectric and polarization results, we observed two pressure-induced magnetic phase transitions around 0.3 and 0.4 GPa that divide the phase diagram into three main regions at low temperatures: FE I, PE II and FE II (Fig. \ref{fig:3}c).      

In the FE I phase (P\textless P$_C$), a primary phase described by \textbf{k}$_1$ = (0, 0, $g$) is always accompanied by a commensurate phase with \textbf{k}$_2$ = (0, 0, 1/4), as seen in the contour maps along the (0 0 L) direction in Fig. \ref{fig:2}a and (2 0 L) direction (Extended Data Fig. S2), indicating a strong competition between them. This is in good agreement with the previous reports \citep{Tian2016, Tian2018}. At ambient pressure, the crystal structure is described by the monoclinic model and the refinement provides the benchmark for the structural refinements under pressure (Extended Data Fig. S3). Considering only the primary phase, we have refined a cycloidal model with spins allowed in any direction with the data collected at 4 K and ambient pressure. The refinement yields a satisfactory result with the cycloidal configuration (Extended Data Fig. S6), giving rise to the magnetic moment 3.75(3) $\mu_B$, close to the previous reports \cite{Rodriguez2015}. The cycloidal spin structure belongs to a polar magnetic point group $21'$ \cite{Rodriguez2015}, which allows the presence of electric polarization along the $a$ axis through the spin-current model \citep{Katsura2005,Mostovoy2005} even though the role of spin-orbit coupling has also been argued due to the presence of strong spin-lattice coupling \citep{Tian2016}.

Around 0.3 GPa, the system evolves into a commensurate magnetic phase featured by a vector \textbf{k}$_3$ = 0 which is associated with a PE phase (PE II) that does not show obvious dielectric and polarization anomaly (Fig. \ref{fig:1}). The magnetic structural refinement (Extended Data Fig. S6) yields the A-type spin structure with all spins along the $a$ axis and antiferromagnetically coupled in both $b$ and $c$ axes (Fig. \ref{fig:3}b). This preference can be attributed to the weak single-ion anisotropy as reported in other Fe-contained materials \citep{Ding}. The refined magnetic moment at 4 K is 3.8(3)$\mu_B$. Such a spin arrangement, according to its magnetic symmetry $P2_1/c'$, does not allow ferroelectricity, in good agreement with our polarization results.

\begin{figure}
\centering
\includegraphics[width=0.8\textwidth]{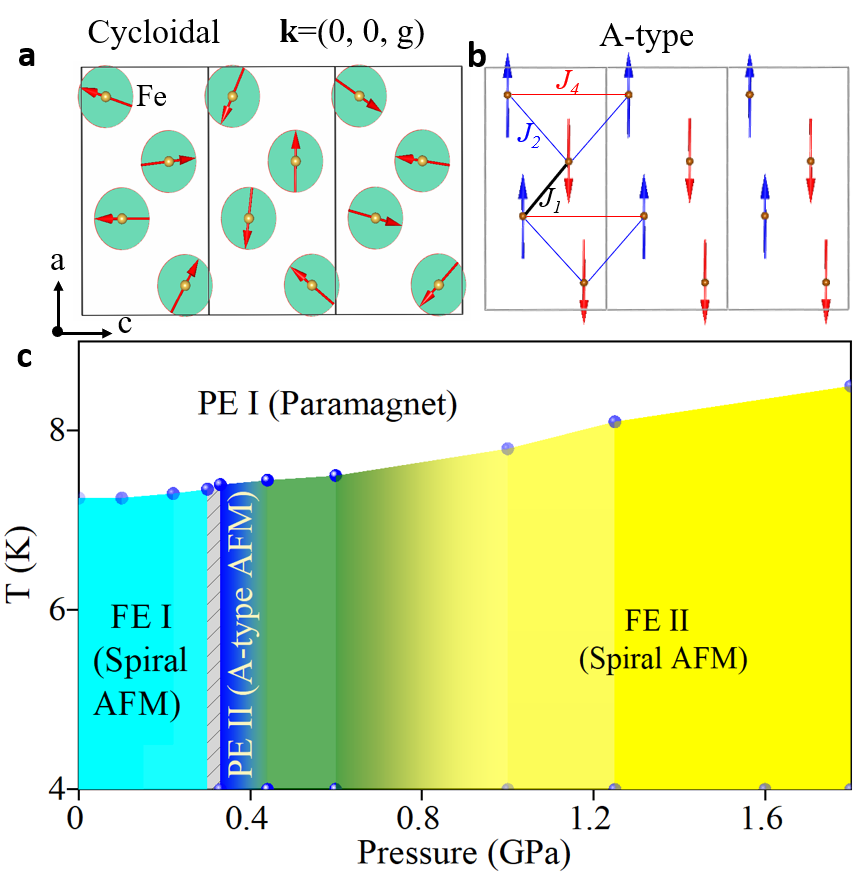}
\caption{\textbf{The schematic drawing of the magnetic structures for the cycloidal and collinear phases, and the magnetic phase diagram of (ND$_4$)$_2$FeCl$_5 \cdot$D$_2$O}. \textbf{a}, The cycloidal spin configuration projected down the $c$ axis. \textbf{b}, The A-type magnetic structure of the collinear phase under the pressure of 0.3 GPa. Only the magnetic Fe ions are shown. \textbf{c}, The magnetic and ferroelectric phase diagram of (ND$_4$)$_2$FeCl$_5 \cdot$D$_2$O. The hatched zone covering a small unknown region in the phase diagram. The regions between the blue and yellow zones stands for the coexistence region of the collinear and the spiral magnetic phases.}\label{fig:3}
\end{figure}

With further increasing pressure, the intensity of magnetic reflection (0 0 1) decreases gradually while the magnetic reflection (0 0 1+$g'$) emerges and develops, eventually forming a region (0.4-0.8 GPa) where the \textbf{k}$_3$ magnetic phase coexists with a new incommensurate phase characterized by \textbf{k}$_4$ = (0, 0, $g'$). Above 0.8 GPa, there is no trace of the \textbf{k}$_3$ magnetic phase and the system transforms into the magnetic ground state consisting of competitive incommensurate (\textbf{k}$_4$) and commensurate phases (\textbf{k}$_2$) (Extended Data Fig. S2). The phase FE II, governed by \textbf{k$_4$} (Fig. \ref{fig:2}) is polar as signified by the electric polarization results in Fig. \ref{fig:1}e. The nuclear reflections at 1 GPa can be well appreciated by the space group $P2_1/a$ corresponding to the monoclinic structure at ambient pressure (Extended Data Fig. S5). The refinement of the cycloidal spin structure based on the magnetic reflections at 1 GPa gives a satisfactory result (Fig. S6 in Extended Data) with the magnetic symmetry identical to that of the phase FE I, meaning the similar ferroelectric properties. This agrees well with the practically linear increment of polarization from the FE I to FE II phase (Fig.\ref{fig:1}e). In the phase FE II, the spin component along the $b$ axis is smallest and the refined angle between the nearest neighboring Fe atoms is roughly half of that in the phase FE I. 

\begin{figure}[t]
\centering
\includegraphics[width=0.8\textwidth]{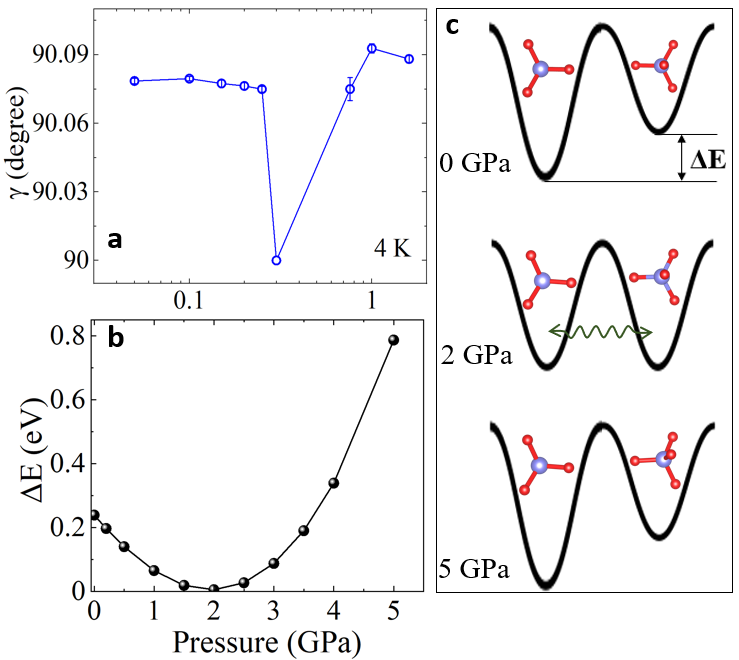}
\caption{\textbf{The modification of pressure on the crystal structures of (ND$_4$)$_2$FeCl$_5 \cdot$D$_2$O and the DFT calculation results.}  \textbf{a}, The lattice parameter $\gamma$ obtained from the peak splitting of the reflection (0 4 0) at 4 K under various pressures. \textbf{b}, The DFT calculated energy difference $\bigtriangleup$E between the two orientations of ND$_4$ as a function of pressure. \textbf{c}, The double-well potential and the associated arrangement of the ND$_4$ group at ambient pressure, 2 GPa and 5 GPa from the DFT calculations. The green line stands for the possible tunneling between the two states.}\label{fig:4}
\end{figure}

To the best of our knowledge, the reentrant spin-driven ferroelectricity (cycloidal spin structure) under pressure has never been documented. To unravel the origin of this novel reentrant phenomenon, we first examine the variation of the crystal structure under pressure. At ambient pressure, (ND$_4$)$_2$FeCl$_5 \cdot$D$_2$O undergoes a structural disorder-order transition at 79 K \citep{Rodriguez2015,Bruning2020}, signaled by the peak splitting on the reflection (0 4 0) (Extended Data Fig. S7). 
Neutron diffraction under pressure at room temperature shows no obvious discontinuity on lattice parameters but they display a larger compression for the $(ac)$ plane compared with the $b$ direction (Extended Data Fig. S8). The peak splitting on the reflection (0 4 0) persists with the pressure below 0.3 GPa (Extended Data Fig. S9 and S10). However, while the pressure approaches P$_C$, the peak splitting becomes negligible coincident with the development of the collinear spin structure. With further increasing pressure, the peak splitting recovers, accompanying with the reentrant ferroelectric phase (FE II). Since the peak splitting is a result of the monoclinic distortion, we can deduce the lattice parameter $\gamma$ as a function of pressure. As shown in Fig. \ref{fig:4}a, while at low and high pressures, $\gamma$ is around 90.08 $^{\circ}$, it becomes 90 $^{\circ}$ at 0.3 GPa. This indicates no monoclinic distortion at 0.3 GPa. Taking the space group $Pnma$ and $P2_1/a$ as structural models, we examined the structural refinements of the neutron data at 0.3 GPa. The monoclinic structure brings more satisfactory results than the orthorhombic model although $\gamma$ becomes 90 $^{\circ}$ at 0.3 GPa (Extended Data Fig. S4). This is consistent with the theoretical calculations in the following.  

To gain further insights into the evolution of the structure under pressure, we carried out density functional theory (DFT) calculations. The spin-polarized DFT calculations examined two sorts of orientations of ammonium ions which are related by the inversion of all the surrounding hydrogen atoms with respect to the N atom at the center of each tetrahedron, forming a double-well potential ((Fig. \ref{fig:4}c)). This potential is featured by the energy difference  $\bigtriangleup$E  between the two orientations of NH$_4^+$. We first confirm that $\gamma$ gradually increases with increasing pressure (Extended Data Table II), indicating that the stable structure at 0.3 GPa is still monoclinic. Furthermore, we evaluated $\bigtriangleup$E as a function of pressure. As shown in Fig. \ref{fig:4}b,  $\bigtriangleup$E first decreases monotonically and then reaches a minimum at 2 GPa. Above 2 GPa, it increases greatly with the evolution of the pressure. The fact that $\bigtriangleup$E becomes negligible at 2 GPa, forming a degenerate state, is associated with the robust permutation between the two orientational NH$_4^+$ cations. This is reminiscent of the quantum tunneling behavior that usually occurs in NH$_4^+$-contained chlorides \citep{Prager1997,Prager1983}. Since the DFT structural optimization ignores the zero-point fluctuation, which could be significant for NH$_4^+$, the pressure required to diminish the energy difference or to promote the quantum tunneling could have been over-estimated. Thus, it becomes possible in (ND$_4$)$_2$FeCl$_5 \cdot$D$_2$O that strong hydrogen bond fluctuations as a result of the rotational tunneling of NH$_4^+$ occur around 0.3 GPa. With further increasing pressure, the double-well potential becomes asymmetric (Fig. \ref{fig:4}c), ruining the quantum tunneling environment, thereby making the system more structurally distorted as found at 1 GPa. As sketched by the ``V" shape in Fig. \ref{fig:1}e and Fig. \ref{fig:4}a, a quantum-critical-like point may exist around 0.3 GPa, implying that PE II phase is practically a quantum paraelectric state.

Even though the magnetic properties of (NH$_4$)$_2$FeCl$_5 \cdot$H$_2$O at ambient pressure are similar to those of TbMnO$_3$, the pressure-dependent magnetism is rather different \cite{Aoyama2014, Terada2016}.  
The evident difference undoubtedly lies in the presence of hydrogen-bond-mediated exchange interactions in (NH$_4$)$_2$FeCl$_5 \cdot$H$_2$O. The importance of the hydrogen bonds has been emphasized recently in metal-organic framework materials \citep{Manson2009} where differing hydrogen-bond motifs in fact can bring distinct magnetic properties. 
In (NH$_4$)$_2$FeCl$_5 \cdot$H$_2$O, four exchanges, $J_1-J_4$, are sufficient to appreciate its magnetic structures (Fig. \ref{fig:1}b) \citep{Campo2008, Rodriguez2015,Clune2019}. 
The inspection of the possible exchange paths in (NH$_4$)$_2$FeCl$_5 \cdot$H$_2$O discloses that the most relevant exchange interaction to the order-disorder of the NH$_4$ is $J_4$ (Fig. \ref{fig:1}a). It can be bridged by the path Fe-Cl···H-N1-H···Cl-Fe involving strong hydrogen bonds (nonlinear Cl···H-N1-H···Cl bond angles: 173.1 $^{\circ}$ and 179.1 $^{\circ}$ and bond lengths: 2.20 \AA~and 2.29 \AA~from the structural model).
When the hydrogen bonds are well oriented, i.e. the situation without pressure, $J_4$ is comparable to $J_2$ in strength \cite{Bai2020}. Therefore, exchanges $J_2$ and $J_4$ form triangular loops and compete when the pressure is smaller than 0.3 GPa, rendering the system strongly frustrated. 
At 0.3 GPa, because of the presence of the rotational tunneling of the ammonium, $J_4$ is expected to be weakened or interrupted, leading to a simple hierarchy of the strength of the exchange couplings $J_1$ $>$ $J_2$ $>$ $J_3$. This leads to the  A-type magnetic configuration that satisfies all the antiferromagnetic interactions. In fact, such the A-type structure was found in other isostructural compounds without NH$_4^+$ groups \citep{Gabas1995}. 
In the reentrant ferroelectric phase, the strong magnetic frustration recovers due to the removal of the bond fluctuations, making the spins to propagate incommensurately. As a result, the magnetism in this molecular material is likely governed by the rotational tunneling of the ammonium groups, which in turn drives the reentrance of the polar phase.
Further experiments such as inelastic neutron scattering will be helpful to ascertain experimentally this tunneling mechanism. 

We have established that magneto-electric properties in (NH$_4$)$_2$FeCl$_5 \cdot$H$_2$O can be rationally controlled in a low-pressure window, thanks to the presence of molecular force couplings and the tunable super-super-exchange interactions. We have shown that the orientations of ammonium cations in the molecular structure, activated by the quantum tunneling, steers the presence and absence of incommensurate magnetic orders which in turn control the spontaneous electric polarization. Such a route might be generalized to other molecular materials, paving ways for the design of multi-functional magnetic materials.

\bibliographystyle{naturemag}

\clearpage

\noindent\textbf{Methods}\\
\noindent\textit{Sample preparation and characterization.} 
Large single crystals of (ND$_4$)$_2$FeCl$_5 \cdot$D$_2$O were grown using aqueous solutions of ND$_4$Cl and FeCl$_3$ with a surplus of HCl for neutron scattering purpose. At the 38$^{\circ}$C growth temperature, optically clear, red single crystals with well-developed flat crystal faces of size above 100 mm$^3$ can be obtained within typically 6 to 8 weeks. The evaporation of the solvent was controlled with a sample environment chamber during the crystal-growth process.  The crystals were characterized with single crystal x-ray diffraction technique. The sample for neutron scattering purpose was deuterated.\\
\noindent\textit{Magnetoelectric property measurement.} 
We have employed a self-clamped piston cylinder cell (PCC) to study pressure effects on the ferroelectric transition of (NH$_4$)$_2$FeCl$_5 \cdot$H$_2$O single crystal. The relative dielectric constant along the orthorhombic a axis under various pressures was measured with an Andeen-Hagerling AH2700 capacitance bridge at a fixed frequency of 1 kHz in a cryogen-free superconducting magnet system (Oxford Instruments, Teslatron PT). A total of three samples labeled as $\#$1, $\#$2, and $\#$3 have been measured under high pressure in this study. Daphne 7373 was used as the pressure transmitting medium. The pressure inside the PCC was determined at room temperature from the relative resistance change of manganin gauge via $P$ (GPa) = $\Delta R/R_0/0.00251$. The temperature dependence of the electric polarization was calculated by time-integrating the pyroelectric current. In order to ensure a single-domain state, a static electric poling field of 100 V mm$^{-1}$ was applied well above the ferroelectric transition temperature. After cooling the crystal to base temperature, the poling field was removed and the pyroelectric current was recorded during subsequent heating process with constant heating rate of 1 K min$^{-1}$.\\
\noindent\textit{Magnetization measurement.}
Temperature dependence of magnetization $M(T)$ under high pressure was measured with a miniature piston-cylinder cell in a commercial magnetic property measurement system (MPMS-III) from Quantum Design. The (NH$_4$)$_2$FeCl$_5 \cdot$H$_2$O single crystals together with a piece of Pb was loaded into Teflon capsule filled with Daphne 7373 as the pressure transmitting medium. The magnetic field was applied along the orthorhombic c axis. The pressure at low temperatures was determined from the superconducting transition of Pb.\\
\noindent\textit{Neutron diffraction.}
Single crystal neutron diffraction experiments were carried out on the fixed-incident-energy triple axis spectrometer HB-1A and HB-3A four-circle diffractometer \citep{HB3A} at the High Flux Isotope Reactor (HFIR), Oak Ridge National laboratory (ORNL). At HB-1A, a monochromatic neutron beam of wavelength $\lambda$ = 2.358 \AA~(E$_i$=14.6 meV) was employed through a double pyrolytic graphite monochromator system.  Highly oriented pyrolytic graphite filters (HOPG) were placed after each monochromator to significantly reduce the higher order contamination of the incident beam. The pressure below 0.3 GPa was applied with a Harwood gas pressure cell at HB-1A within $ac$ plane. The above 0.3 to 1.8 GPa pressure range data were collected at HB-3A with a CuBe pressure cell (Daphne oil 7373 as the pressure transmitted medium) using $\lambda$ = 1.546 \AA~ neutron beam. The large scale pressure effect on the magnetic phase is studied with neutron scattering techniques. We measured the (2 0 L) and (0 0 L) planes in the low pressure range of 0-0.2 GPa with a Hardwood gas pressure cell. The higher pressure range for (2 0 L), (0 0 L) and (0 1 L) planes was carefully investigated. Combining both low and high pressure range of the (0 0 L), one can have basic information to construct a magnetic vs pressure phase diagram. The results of the (2 0 L) and (0 1 L) planes contain more information due to finer resolution. The magnetic structures under different pressures were determined by combining symmetry analysis and refinements using the Bilbao Crystallographic Server (Magnetic Symmetry and Applications software \citep{Bilbao} and Fullprof suite program \citep{Fullprof}.\\ 
\noindent\textit{Spin-polarized Density Functional Theory calculations.}
Spin-polarized Density Functional Theory (DFT) calculations were performed using the Vienna Ab initio Simulation Package (VASP) \citep{Kresse1996}. The calculation used Projector Augmented Wave (PAW) method \citep{Blochl1994,Kresse1999} to describe the effects of core electrons, and Perdew-Burke-Ernzerhof (PBE) \citep{Perdew1996} implementation of the Generalized Gradient Approximation (GGA) for the exchange-correlation functional. The lattice parameters and atomic coordinates determined by neutron diffraction at 2 K \citep{Rodriguez2015} were used as the initial structure, with an anti-ferromagnetic spin configuration as shown in Fig. 3b. For unit cells under pressure, the initial lattice constants were calculated by using the ambient pressure values and the equation of state as determined from our pressure experiment. The electronic structure was calculated on a 2$\times 3\times4~\Gamma$-centered mesh for the unit cell. The total energy tolerance for electronic energy minimization was 10$^{-8}$ eV. Structural relaxations were performed with fixed unit cell volume for the original configurations ($\gamma >$ 90$^{\circ}$) at various pressures. All atoms and the cell shape are relaxed to the local potential energy minimum. The $\gamma$ angle is seen to increase with increasing pressure. From these optimized monoclinic configurations, all ammonium ions are flipped as depicted in Fig. 4. After flipping the ammonium, the relaxation procedures was carried out. In the calculation, the flipped ammonium ions are relaxed while all other atoms and the cell are fixed, so that $\bigtriangleup$E can be calculated. The energy tolerance for structure optimization was 10$^{-7}$ eV, and the maximum interatomic force after relaxation was below 0.001 eV/\AA. A Hubbard U term of 4.0 eV was applied on Fe for the 3d electrons \citep{Wang2006}. The optB86b-vdW functional \citep{Klimes2010} for dispersion corrections was used to describe the van der Waals interactions. \\
\noindent\textbf{Acknowledgments}\\
The research at Oak Ridge National Laboratory (ORNL) was supported by the U.S. Department of Energy (DOE), Office of Science, Office of Basic Energy Sciences, Early Career Research Program Award KC0402010, under Contract DE-AC05-00OR22725. This research used resources at the High Flux Isotope Reactor, a DOE Office of Science User Facility operated by ORNL. JGC was supported by the National Key R \& D Program of China (Grant No. 2018YFA0305700), the National Natural Science Foundation of China (Grants No. 11921004, 11874400, No. 11834016), the Beijing Natural Science Foundation (Grant No. Z190008), the Strategic Priority Research Program and Key Research Program of Frontier Sciences of the Chinese Academy of Sciences (Grants No. XDB25000000 and No. QYZDB-SSW-SLH013) as well as the CAS Interdisciplinary Innovation Team. This manuscript has been authored by UT-Battelle, LLC under
Contract No. DE-AC05-00OR22725 with the U.S. Department of Energy. The United States Government
retains and the publisher, by accepting the article for publication, acknowledges that the United States
Government retains a non-exclusive, paidup, irrevocable, world-wide license to publish or reproduce the
published form of this manuscript, or allow others to do so, for United States Government purposes. The
Department of Energy will provide public access to these results of federally sponsored research in
accordance with the DOE Public Access Plan (http://energy.gov/downloads/doepublic-access-plan).
\\
\noindent\textbf{Author contributions}\\
H. C. supervised this work.  H. C. and W. T. conceived the project. W. T. synthesized the crystals and B. C. C. carried out the single crystal x-ray diffraction and identified the crystal structures at high and low temperatures. Y. W., L. D. and H. C. performed neutron experiments and analyzed the data. N. S., Y. M., K. Z., Y. S. and J. C. performed the magnetization and electric polarization measurements under pressure through discussion with H. C.. Y. Q. C. performed the Density Functional Theory calculations with inputs from H. C., L. D., Y. W., W. T. and X. B.. L. D., Y. W. and H. C. wrote the paper with comments from all the authors. Y. W., L. D. and N. S. contributed equally to this work.\\

\noindent\textbf{Competing interests}\\
The authors declare no competing interests.

\end{document}